\documentclass[twocolumn,showpacs,preprintnumbers,amsmath,amssymb]{revtex4}
\usepackage{graphicx}% Include figure files
\usepackage{epsfig}
\usepackage{dcolumn}% Align table columns on decimal point
\usepackage{bm}% bold math
\begin{document}
\preprint{}
%%%%%%%%%%%%%%%%%%%%%%%%%%%%%%%%%%%%%%%%%%%%%%%%%%%%%%%%%%%%%%%%%%%%%%%%%%%%%%
\title{Field dependence of the vortex core size in a multi-band superconductor}
\author{F.D.~Callaghan$^{1}$, M.~Laulajainen$^{1}$, C.V.~Kaiser$^{1}$, and J.E.~Sonier$^{1,2}$}
\affiliation{$^1$Department of Physics, Simon Fraser University, Burnaby, British Columbia V5A 1S6, Canada \\
$^2$Canadian Institute for Advanced Research, 180 Dundas Street West, Toronto, Ontario M5G 1Z8, Canada}
\date{\today}
%%%%%%%%%%%%%%%%%%%%%%%%%%%%%%%%%%%%%%%%%%%%%%%%%%%%%%%%%%%%%%%%%%%%%%%%%%%%%%%
\begin{abstract}
The magnetic field dependence of the vortex core size in the multi-band superconductor NbSe$_2$ has been determined
from muon spin rotation measurements.  The spatially extended nature of the 
quasiparticle core states associated with the smaller gap leads to a rapid field-induced shrinkage of the core size 
at low fields, while the more tightly bound nature of the states associated with the larger gap leads to a 
field-independent core size for fields greater than 4 kOe.
A simple model is proposed for 
the density of delocalized core states that establishes a direct relationship between the field-induced reduction 
of the vortex core size and the corresponding enhancement of the electronic thermal conductivity. We show that this model
accurately describes both NbSe$_2$ and the single-band superconductor V$_3$Si.  
\end{abstract}
\pacs{74.25.Ha, 74.25.Jb, 74.25.Qt, 74.70.Ad, 76.75.+i}
\maketitle
%%%%%%%%%%%%%%%%%%%%%%%%%%%%%%%%%%%%%%%%%%%%%%%%%%%%%%%%%%%%%%%%%%%%%%%%%%%%%%%

The discovery of MgB$_2$ \cite{Nagamatsu}, a seemingly conventional type-II superconductor with a high 
transition temperature $T_c \! = \! 39$~K, has sparked considerable experimental and theoretical interest 
in the physics of multi-band superconductivity (MBSC). 
MBSC, which  was first treated theoretically in 1959 using BCS theory \cite{Suhl}, implies that below $T_c$ 
distinct energy gaps open up on different 
sheets of the Fermi surface.  In MgB$_2$, a second smaller gap is induced on the $\pi$ bands through coupling 
to the intrinsically superconducting $\sigma$ bands.  While the high value of $T_c$ 
and the large boron isotope effect in MgB$_2$ are now believed to originate from strong electron-phonon 
coupling on the two-dimensional $\sigma$ bands, it is also clear that several properties of
MgB$_2$ are a manifestation of MBSC \cite{Eskildsen,Gonnelli,Cubitt,Tsuda,Bugoslavsky}.
Large vortex cores have been observed on the $\pi$ bands by scanning 
tunneling spectroscopy \cite{Eskildsen}. 
Microscopic calculations of the local density of states (DOS) in a two band
superconductor predict that the smaller gap will give rise to  
spatially extended quasiparticle (QP) states at low field \cite{Nakai:02,Ichioka:04}.  At higher 
field, these loosely bound states become delocalized, with the core size now being determined by
the more localized states associated with the larger gap.
This model is supported by the magnetic field dependence of the
electronic thermal conductivity \cite{Sologubenko,Kusunose} and the specific heat \cite{Yang,Bouquet:02,Tewordt}, 
techniques which are sensitive to the delocalization of QPs. 

Muon spin rotation ($\mu$SR) 
has proven to be an accurate method for determining the field dependence of the vortex core size 
\cite{Sonier:04b}, however single crystals are required for this kind of measurement. Although there 
have been several studies of MgB$_2$ in the vortex state by $\mu$SR 
\cite{Panagopoulos,Niedermayer,Ohishi,Serventi,Angst}, 
these have been done on powdered samples because of the difficulty of growing large, clean, MgB$_2$ 
single crystals.

NbSe$_2$ has long been thought of as a typical $s$-wave type-II superconductor. However, there is now
convincing evidence that it too is a multi-band superconductor \cite{Corcoran,Yokoya,Boaknin:03,Rodrigo}.
Experiments suggest that an energy gap of magnitude $\sim$ 1 meV exists on two bands derived from Nb $4d$ 
orbitals, and a gap $\sim$ 2-3 times smaller also exists on a band derived from Se $4p$ orbitals. 
In the quest to fully understand the physics of MBSC, it is important to study superconductors other than
MgB$_2$, and to this end we performed $\mu$SR measurements of the field dependence of the vortex core size in NbSe$_2$ 
at very low temperature.  
A distinct advantage of NbSe$_2$ is the availability of large, high quality, 
single crystals.  Single crystals allow for the alignment of the initial muon spin polarization along
a single principal crystallographic axis.  In this case the $\mu$SR lineshape resembles the asymmetric theoretical
internal magnetic field distribution $n(B)$ of the vortex lattice.  In particular, the high-field tail 
is the contribution of the vortex cores to $n(B)$ (see Fig.~1) \cite{Sonier:00}.
Previous $\mu$SR measurements of the field-dependence of the core size in NbSe$_2$ were performed at
$T \! \geq \! 2.3$~K \cite{Sonier:97a}.  At such temperatures, thermal excitations of the bound core states 
(the so-called `Kramer-Pesch effect' \cite{Kramer,Miller}) result in significant overlap of the QP wavefunctions of 
adjacent vortices.  When this happens the QPs delocalize, giving rise to a strongly field-dependent core size 
\cite{Sonier:04a}, as predicted by the microscopic theory \cite{Ichioka:99}.

In order to freeze out thermal 
excitations of the bound core states and fully isolate the effect of magnetic field on the core size, we have
carried out $\mu$SR measurements on a single crystal of NbSe$_2$ ($T_c =$ 7.0 K and upper critical 
field $H_{c2} =$ 45 kOe) in a dilution refrigerator at $T \! = \! 20$~mK.  
This also permits a direct comparison with electronic thermal conductivity $\kappa_e$ data, 
which is a measure of the extent of QP delocalization.
The experiment was performed at the Tri-University Meson Facility (TRIUMF), Vancouver, 
Canada, with  the magnetic field applied parallel to the crystallographic [001] axis.
For each measurement, the field was applied at $T > T_c$ before cooling the sample to the 
desired temperature. 

The spin-polarized positive muons implanted into the sample stop randomly on the length scale of the 
vortex lattice, and therefore uniformly sample $n(B)$.  
Each muon precesses around the local field $B$ at its site at the Larmor frequency 
$\omega = \gamma_{\mu}B$, where $\gamma_{\mu}/2\pi$ = 135.5342 MHz/T is the muon gyromagnetic ratio.  
On decay of the muon after an average lifetime of 2.2 $\mu$s, a positron is emitted preferentially
along the direction of the muon spin.  The time evolution of the muon spin polarization is determined by
detecting decay positrons from an ensemble of $\sim 2 \times 10^7$ muons.
The exact functional form of the muon spin polarization depends on $n(B)$.  
Further details of the experimental technique used here can be found in Ref.~\cite{Sonier:00}.
The $\mu$SR time spectra were fit assuming a Ginzburg-Landau (GL) model for the spatial field profile
given by
\begin{equation}
B({\bf r}) = B_0 (1-b^4) \sum_{{\bf G}}
\frac{e^{-i {\bf G} \cdot {\bf r}} \, \, u \, K_1(u)}{\lambda_{ab}^2 G^2} \,,
\label{eq:field}
\end{equation}   
where $b = B/B_{c2}$, $B_0$ is the average internal field, {\bf G} are the reciprocal lattice vectors,
$K_1(u)$ is a modified Bessel function, $u^2 \! = \! 2 \xi_{ab}^2 G^2 (1 + b^4)[1-2b(1 - b)^2]$,
$\xi_{ab}$ is the GL coherence length, and $\lambda_{ab}$ is the magnetic penetration depth.  As explained in 
Ref.~\cite{Sonier:04b}, previous $\mu$SR works have demonstrated that the field-dependence of the parameter 
$\xi_{ab}$ reflects changes in the vortex core size due to changes in the electronic structure of the vortex cores.
\begin{figure}
\centerline{\epsfxsize=3.4in\epsfbox{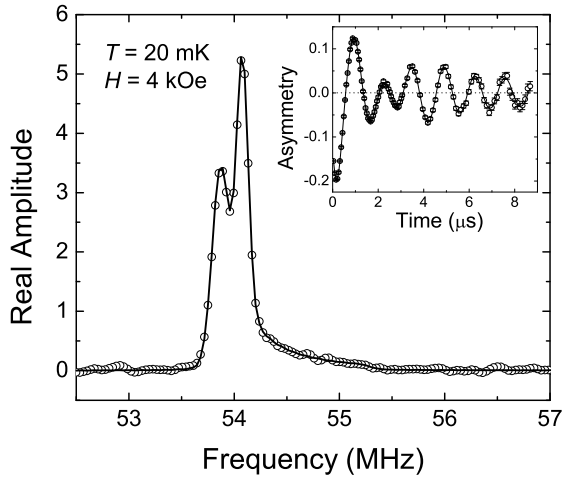}}
\caption{Fast Fourier transforms of both the time dependence of the muon spin polarization in
NbSe$_2$ at $T =$ 20 mK and $H =$ 4 kOe (open circles) and the fit (solid curve) to Eq.~(1).
Inset: muon spin precession signal and fit to Eq.~(1) (in a reference frame rotating at 53.3 MHz).}
\end{figure}

Figure~1 shows fast Fourier transforms (FFTs) of both the muon spin precession signal at $H =$ 4 kOe and the 
fit to Eq.~(\ref{eq:field}).  The FFTs closely resemble $n(B)$ for a hexagonal vortex lattice, 
but are significantly broadened by the apodization used to reduce ringing and noise in the FFT due to the 
finite time-range and the short muon lifetime, respectively.  
The lineshape is also broadened by a Gaussian distribution of fields from randomly oriented nuclear dipolar 
moments.  The width of this distribution was determined to be $\approx 0.25$~MHz.
In addition, there is a background peak at $\sim$~54.1~MHz originating from  muons stopping outside the sample.

Recently, $\mu$SR measurements on powder samples of Mg$_{1-x}$Al$_x$B$_2$ were modeled assuming 
two distinct energy gaps, and a two-component field distribution for the vortex lattice \cite{Serventi}.
Specifically, this model assumes two distinct coherence lengths, which implies that both large and small
vortex cores exist simultaneously in the sample.
However, for NbSe$_2$ we find that the data are well described by Eq.~(\ref{eq:field}).
A visual inspection of Fig.~1 shows that the fit captures all of the main features of the $\mu$SR lineshape.
This is consistent with recent theoretical work predicting that the formation of two vortex sublattices is 
energetically forbidden in a two-gap superconductor \cite{Babaev}.
\begin{figure}
\centerline{\epsfxsize=3.4in\epsfbox{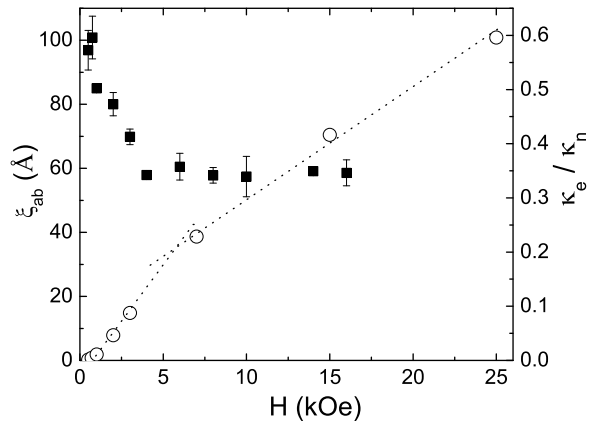}}
\caption{Field dependence of the vortex core size $\xi_{ab}$ at $T =$ 20 mK (squares) and the
electronic thermal conductivity $\kappa_e$ normalised to its normal-state value $\kappa_n$ (circles).
The dashed lines are guides for the eye.}
\end{figure}

Recent $\mu$SR measurements of the field dependence of the core size in the single-band superconductor
V$_3$Si \cite{Sonier:04a} have confirmed theoretical predictions of field-induced core shrinkage due to QP 
delocalization \cite{Ichioka:99}.  In the case of NbSe$_2$, the delocalization of core states 
is dependent on both energy gaps.  At low fields the rapid reduction of $\xi_{ab}$ and the simultaneous
rapid increase of $\kappa_e$ with increasing magnetic field are consistent with this regime being dominated by  
loosely bound states associated with the smaller energy gap \cite {Nakai:02,Ichioka:04}.  If the loosely bound
core states associated with the smaller gap have completely delocalized by $H \approx$ 4 kOe, then we
expect the behavior of $\xi_{ab}$ at larger fields to be dominated by the larger gap.
The saturation of $\xi_{ab}$ above 4 kOe indicates that the QPs are highly localized in the vortex cores
at higher fields.  
This is in line with STM measurements which show that the bound core states at $H = 10$~kOe are more spatially
confined than those at $H = 1$~kOe \cite{Hess}.
We note that even though the shrinking of the core size saturates above 4 kOe, $\kappa_e$ still exhibits a 
considerable dependence on field.
The reason is that although the number of 
delocalized QPs per vortex is no longer changing, the increasing number of vortices increases the total number 
of delocalized QPs (i.e. heat carriers) in the sample.

Given that the core size at low (high) field is primarily determined by the small (large) gap, one might naively
expect that the BCS relationship $\xi \sim v_F/\pi\Delta_0$, where $v_F$ is the Fermi velocity and $\Delta_0$ 
is the magnititude of the energy gap, could be used to 
determine the ratio of the high and low field core sizes.  However, recent calculations based on GL theory 
have shown that this relationship is not applicable to a multi-band superconductor and that a simple expression 
for the ratio of the two core sizes only exists in the limit of zero interband coupling \cite{Zhitomirsky}.  
For finite interband coupling, the ratio must be calculated numerically and is dependent on a number of material 
parameters.

\begin{figure}
\centerline{\epsfxsize=3.4in\epsfbox{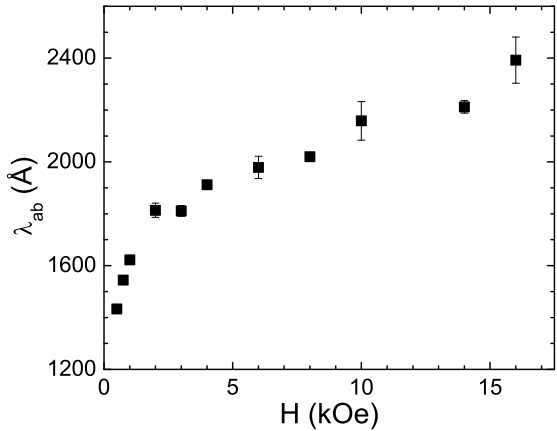}}
\caption{Field dependence of the magnetic penetration depth in NbSe$_2$ at $T = 20$~mK.}
\end{figure}
As can be seen from Eq.~(1), the magnetic penetration depth $\lambda_{ab}$ is also a fitting parameter in our
analysis, extracted simultaneously with $\xi_{ab}$.  
Fig.~3 shows that $\lambda_{ab}$ increases strongly with 
$H$ at low fields and displays a weaker, though significant, dependence at higher fields.
The Volovik effect \cite{Volovik} has previously 
been invoked to explain the field dependence of $\lambda_{ab}$ in $s$-wave type-II superconductors \cite{Kadono}.  
This effect involves a 
shift of the quasiparticle energy spectrum by an energy $\sim {\bf v_F} \cdot {\bf v_S}$, where ${\bf v_S}$ is 
the supercurrent velocity, and it plays a significant role in superconductors with gap nodes.
However, for an isotropic $s$-wave energy gap, a significant dependence of $\lambda_{ab}$ on $H$ is only expected if 
the thermal energy $k_BT$ (where $k_B$ is Boltzmann's constant) is comparable to the magnitude of the energy gap.  
At $T = 20$~mK, $k_BT \approx 10^{-3}$~meV,
whereas the size of the small energy gap in NbSe$_2$ is $\approx 10^{-1}$~meV.  Therefore we cannot attribute the
observed behaviour of $\lambda(H)$ to the Volovik effect.

Calculations of the electronic structure of a single vortex using the Bogoliubov-deGennes equations 
have shown that the presence of extended quasiparticle states significantly modifies the current density, 
and hence $B({\bf r})$, around the vortex core \cite{Gygi}.  With increasing magnetic field these modified 
regions overlap.  Since Eq.~(1) does not account for this, the effect on $B({\bf r})$ shows up in our measurements 
as a field dependent $\lambda_{ab}$.  
In other words $\lambda_{ab}$ in Eq.~(1) is an {\it effective} magnetic penetration depth.
We note that the field dependence of $\lambda_{ab}$ is stronger at low field due to the more rapid delocalization of 
QPs from the vortex cores.

Next we propose a simple model which directly relates our measurements of the vortex core size to the
measurements of the electronic thermal conductivity.
In the conventional picture of a type-II superconductor, the density of {\it localized} QP states 
at the Fermi energy $N(E_F)_{\rm loc}$ is proportional to $\pi\xi^2H$, where $\pi\xi^2$ is the area of a single
vortex core and the density of vortices in the sample is proportional to $H$ \cite{Caroli,Fetter}.  
If we assume that the field-induced density of {\it delocalized} QP states is equal to the reduction in 
$N(E_F)_{\rm loc}$, then    
\begin{equation}
N(E_F)_{\rm deloc} \propto (\pi\xi_0^2 - \pi\xi^2) H,
\label{eq:ne}
\end{equation}
where $\xi_0 \approx 97$~{\AA} is the low-field value of the core size.  Note that
$(\pi\xi_0^2 - \pi\xi^2)H$ is also proportional to the reduction of the total core area in the sample
due to the delocalization of QP core states.

In a metal, $\kappa_e = (1/3)Cv_Fl = (1/9)\pi^2N(E_F)v_Flk_B^2T$, where $C$ is the heat capacity per unit volume,
$k_B$ is Boltzmann's constant, $v_F$ is the Fermi velocity, and $l$ is the electron mean free path \cite{Kittel}.  
Since it is the delocalized QPs that carry heat then $\kappa_e \propto N(E_F)_{\rm deloc}$, and it follows from 
Eq.~(2) that
\begin{equation}
\kappa_e \propto (\pi\xi_0^2 - \pi\xi^2) H.
\label{eq:kappa}
\end{equation}
\begin{figure}
\centerline{\epsfxsize=3.4in\epsfbox{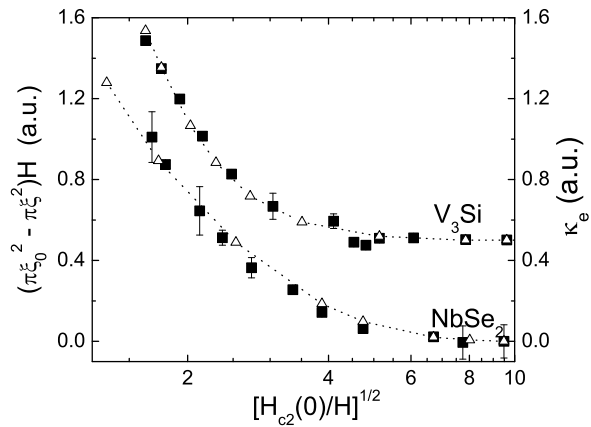}}
\caption{Total reduction in core area (left axis, solid squares) and electronic thermal conductivity 
$\kappa_e$ (right axis, open triangles) plotted against $(H_{c2}/H)^{1/2}$ (which is proportional to the
intervortex spacing) for V$_3$Si and NbSe$_2$.  Both quantities are
normalised to their values at $H \sim$~70~kOe for V$_3$Si and $H \sim$~10~kOe for NbSe$_2$.  
The dashed lines connect the $\kappa_e$ data points.
(The V$_3$Si data are shifted upwards by 0.5 on the vertical scale for clarity.)}
\end{figure}
To show that this relationship is physically valid, in Fig.~4 we plot the measured quantities $\kappa_e$ and
$(\pi\xi_0^2~-~\pi\xi^2)H$ against $(H_{c2}(0)/H)^{1/2}$, which is proprtional to the intervortex spacing.  
We also plot our earlier measurements on V$_3$Si (Ref.~\cite{Sonier:04a}) in this way.
It can be seen that at large intervortex spacing
(low $H$) neither quantity changes significantly.  However, as the field is increased and the vortices
are brought closer together, both quantities exhibit the same field dependence within experimental uncertainty.  
This lends further support to the conclusion that in both single-band V$_3$Si 
and multi-band NbSe$_2$ the field dependences of $\xi_{ab}$ and $\kappa_e$ have a common underlying cause, 
namely the delocalization of bound QPs.

In summary, we have measured the field dependence of the vortex core size in the multi-band
superconductor NbSe$_2$ at $T = 20$~mK.  The observed field dependence is explained by the effects of 
two superconducting energy gaps on the bound core states.
In addition, we have experimentally established a direct
correlation between the field-induced reduction of the vortex core area and the electronic
thermal conductivity in both single-band and multi-band superconductors.

This work was supported by the Natural Sciences and Engineering Research Council (NSERC) of Canada
and the Canadian Institute for Advanced Research (CIAR).  The authors wish to thank the TRIUMF support staff 
for technical assistance, Roger Miller for useful discussions and assistance with data acquisition, 
and J.W. Brill for providing the NbSe$_2$ sample.

%%%%%%%%%%%%%%%%%%%%%%%%%%%%%%%%%%%%%%%%%%%%%%%%%%%%%%%%%%%%%%%%%%%%%%%%%%%%

% REFERENCES

\end{document}